\documentclass[%
 reprint,
 amsmath,amssymb,
 aps,
 rmb,
 longbibliography,
 lengthcheck,%
]{revtex4-1}

\usepackage{graphicx}
\usepackage{dcolumn}
\usepackage{bm}
\usepackage{hyperref}

\usepackage[T1]{fontenc}
\usepackage[latin9]{inputenc}

\usepackage{amsmath}
\usepackage{graphicx}
\usepackage[english]{babel}
\makeatletter

\begin{document}

\title{Thermalization of Interacting Fermions and Delocalization in Fock space }

\author{Clemens Neuenhahn and Florian Marquardt}

\affiliation{Department of Physics, Arnold Sommerfeld Center for Theoretical Physics, and Center for NanoScience\\ 
Ludwig-Maximilians-Universit\"at M\"unchen, Theresienstr. 37, 80333 Munich, Germany
}

\affiliation{Friedrich-Alexander-Universit\"at Erlangen-N\"urnberg\\Institute for Theoretical Physics II, Staudtstr.7, 91058 Erlangen, Germany}

\begin{abstract}
By means of exact diagonalization, we investigate the onset of 'eigenstate
thermalization' and the crossover to ergodicity in a system
of 1D fermions with increasing interaction.
We show that the fluctuations in the expectation values of the momentum
distribution from eigenstate to eigenstate decrease with increasing
coupling strength and system size. It turns out that these fluctuations
are proportional to the inverse participation ratio of eigenstates
represented in the Fock basis. We demonstrate that eigenstate thermalization should set in 
even for vanishingly small perturbations in the thermodynamic limit.
\end{abstract}

\maketitle


\emph{Introduction}. \textendash{} 
Statistical physics relies on the assumption that the system 
under investigation is in thermal equilibrium.  
However,  
what are the precise conditions for an isolated system to relax to thermal equilibrium? 
This question has a long history including the ground breaking numerical
 experiments initiated by Fermi, Pasta and Ulam~\cite{1955_Fermi} on an anharmonic 
 chain of classical oscillators, where thermalization was not observed as expected~\cite{1992_Ford_Fermi_Pasta_Ulam}. 
 Nowadays, the investigation of 
 thermalization in quantum many-body systems attracts a lot of theoretical attention, inspired 
 by the new experimental possibilities in systems of cold atoms ~\cite{2002_GReinerCollapseRevival, 2006_Kinoshita,2007_Hofferbeth}. 
  
The trajectory of a classical ergodic system reaches all regions on the energy shell for sufficiently
long times, establishing the microcanonical ensemble. As a consequence, suitable chosen
subsystems obey the Boltzmann distribution. In the quantum case, switching on an interaction in
a many-body system will combine the unperturbed eigenstates $\left | i \right\rangle$ of similar energies into new
energy eigenstates: $|\alpha\rangle=\sum_{i} c_{i}^{\alpha}\left|i\right\rangle$. If the expectation values $A_{\alpha}=\langle \alpha| \hat{A}|\alpha\rangle$ of observables in these new eigenstates
approach their microcanonical values $A_{{\rm micro}}(E)$, as obtained by averaging over all unperturbed states
in a small energy window around $E$, then the properties of thermal equilibrium are established in each
many-body eigenstate. 
This is the essential idea behind the 'eigenstate 
thermalization hypothesis' (ETH)~\cite{1991_Deutsch,1994_Srednicki}. 

Recently, the ETH has been tested in
numerical experiments~\cite{2008_RigolNature,2009_RigolBreakdown,2009_RigolFermions}, 
by means of exact diagonalization. 
For few-body observables like the momentum distribution, indeed it
was demonstrated that $A_{\alpha}\approx A_{{\rm micro}}(E_{\alpha})$
and that the fluctuations around $A_{{\rm micro}}$ decrease with increasing
interaction strength and system size. 

In the present work, we address the important question of how fast
thermal equilibrium is approached when increasing the system size.
A direct, brute-force numerical approach would be prohibitive. Instead,
we characterize the gradual delocalization of eigenstates in the many-body Fock space
via the inverse participation ratio (IPR)   $\sum_{i=1}^{D}(p_{i}^{\alpha})^{2}$ (with $p_i^{\alpha}=|c_i^{\alpha}|^2$) which turns out
to be connected with the fluctuations of $A_{\alpha}$. While a connection between the IPR and the fluctuations was observed recently~\cite{2010_Santos,2010_Canovi},
 we are able to conjecture its functional form and its dependence on system size and interaction strength, based on earlier
analytical results on Fock-space localization by P.~Silvestrov. In particular,  we have numerical evidence that the interaction strength needed for thermalization
is below that needed for full quantum chaos. Moreover, we find that in the thermodynamic limit (TDL) thermalization (in the sense of the ETH) sets in for arbitrarily small interactions.
This is in contrast to recent observations on relaxation in a classical 1D system~\cite{2009_Cassidy}.  

\begin{figure}
\includegraphics[width=1\columnwidth]{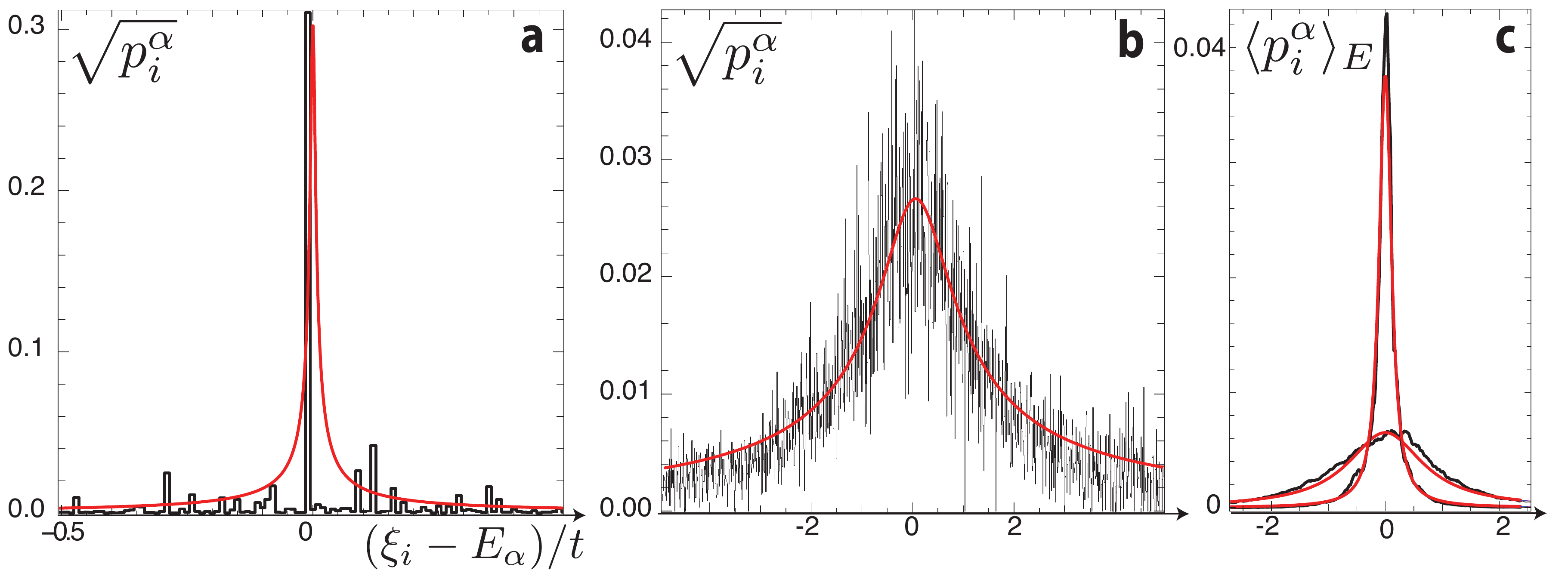}

\caption{Probability distribution $p_i^{\alpha}=|\langle i|\alpha\rangle|^2$ of a many-body eigenstate $\left|\alpha\right\rangle$  in the non-interacting Fock basis (energies $\xi_{i}$). a) For weak interaction $V/t=0.1$, the eigenstate is localized in Fock space, consisting of a few isolated peaks. b) At large $V/t=1.3$, all Fock states with energies $\xi_{i}$ close to $E_{\alpha}$ contribute. c)  $p_i^{\alpha}$ averaged over  a couple of nearby eigenstates in a range $\delta E/t=0.05$, for $V/t=0.45,1.45$ (top, bottom). It can be approximated by a Lorentzian of width $\overline{\Gamma}$ (red line). The energy was chosen to correspond to infinite effective temperature (see main text).}
\label{Fig1}
\end{figure}

Here, we address these questions by 
means of exact numerical diagonalization for a system of spinless 1D fermions on a 
lattice, where integrability is broken by an interaction of strength $V$. As the observable of interest, we consider the fermionic momentum distribution $\hat{f}_{k}$. Our main result is that for large enough $V$ the fluctuations of
$f_k^{\alpha}\equiv\langle\alpha|\hat{f}_{k}|\alpha\rangle$ are determined by 
the IPR  which
roughly can be considered as the inverse number of non-interacting
Fock states $|i\rangle$ contributing to $|\alpha\rangle$ (see e.g. ~\cite{2000_SilvestrovNum}).
The IPR itself keeps track of the transition from integrability to
quantum chaos~\cite{1997_Silvestrov,2010_Santos} and it was conjectured only recently that it might directly determine the deviations of steady state expectation values from the corresponding microcanonical value~\cite{2009_Olshanii}. \\
We observe \emph{three different regimes}, depending on the interaction strength. An important scale is set by the mean level
spacing $\Delta_{f}$ between Fock states that couple directly to a given initial Fock state. If the interaction is smaller than  $\Delta_{f}$, then the eigenstates 
are 'localized' in Fock space~\cite{1997_AlthsulerLoc,1997_Georgeot} and 
experience only a perturbative correction due to the interaction (see Fig.$\,$\ref{Fig1}a). For couplings
 beyond $\Delta_{f}$, the eigenstates delocalize and remarkably 
 the IPR decreases \emph{exponentially} with $V$ on a scale that depends on
 $\Delta_{f}$. This scale essentially decreases polynomially in particle number and system size. Therefore, 
we expect the fluctuations of $f_k^{\alpha}$ to be suppressed to zero in the TDL even for vanishingly small interaction strength,
 establishing eigenstate thermalization of the considered observable. Increasing the interaction
  even further, eigenstates become chaotic (see Fig.$\,$\ref{Fig1}b) and the IPR
 as well as the fluctuations in $f_k^{\alpha}$ decrease as the inverse many-body density of states as it 
 was conjectured in ~\cite{1991_Deutsch,1994_Srednicki}. These results should apply rather generically to few-body observables diagonal in the eigenbasis of the unperturbed Hamiltonian.  

\emph{Model}. \textendash{} We consider $n$ spinless 1D fermions with
periodic boundary conditions on a lattice of $N$ sites  
and with a next-nearest neighbor interaction breaking the integrability of the system. The Hamiltonian reads: 
\begin{eqnarray}
\hat{H}_{0}+\hat{V} & = & -t\sum_{i=1}^{N}\hat{c}_{i}^{\dagger}\hat{c}_{i+1}+{\rm H.c.}\nonumber \\
 &  & +V\sum_{i=1}^{N}\left(\hat{n}_{i}-1/2\right)\left(\hat{n}_{i+2}-1/2\right).\label{eq0}\end{eqnarray} 
 The eigenstates $|i\rangle$ of $\hat{H}_0$  with $\xi_i=\langle i|\hat{H}|i\rangle$  are given by the Fock states of $n$ fermions
in momentum space. Due to the translational symmetry,  the interaction does not mix Fock states with different total momentum $K$. Therefore, each momentum sector $K$ with dimension $D_{K}$ will be considered separately.We exclude the $K=0$-sector as it possesses a trivial extra symmetry under reflection. In our numerical examples $n=7$ and $N=21$.

\begin{figure}
\includegraphics[width=1\columnwidth]{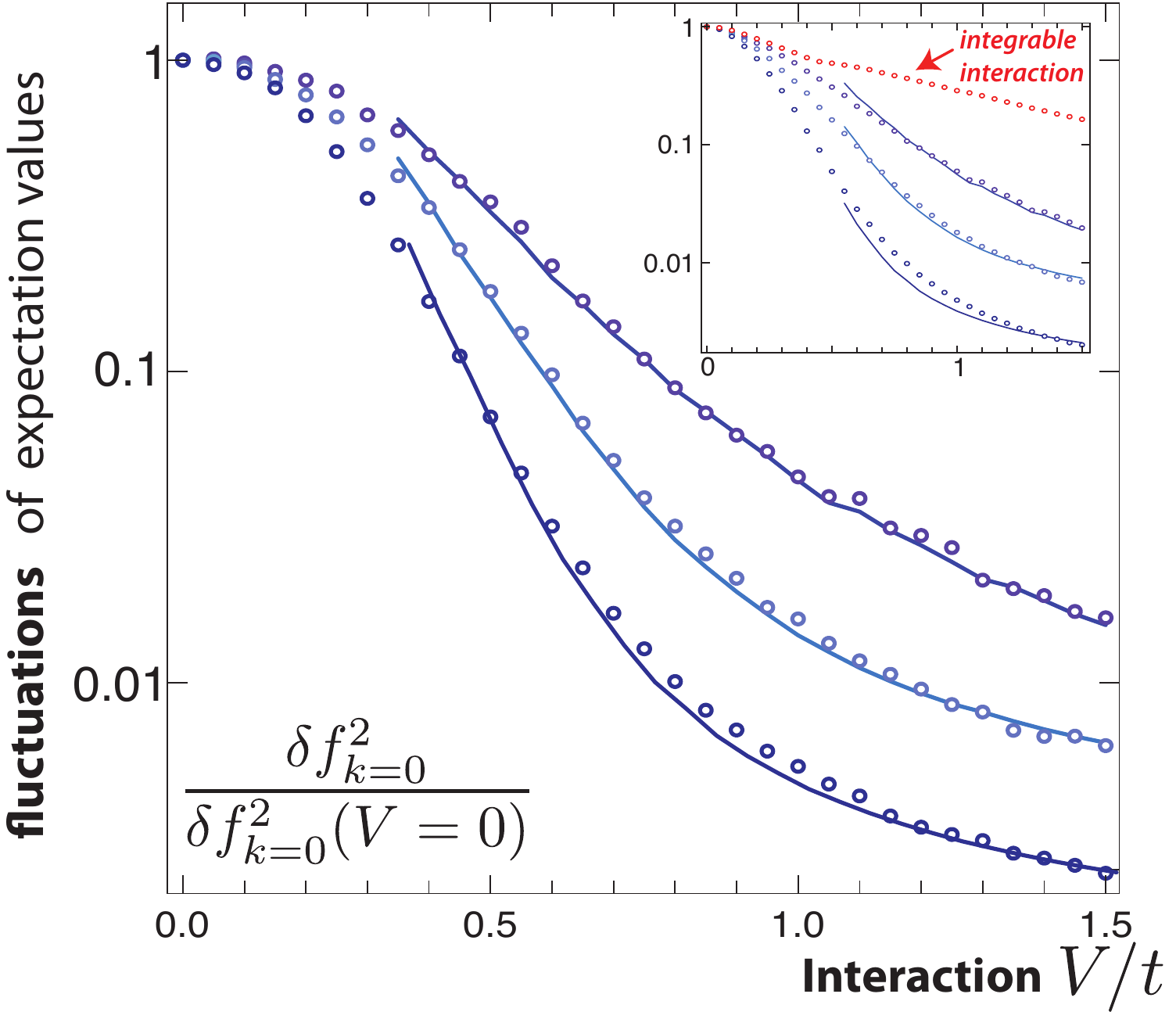}
\caption{Fluctuations of expected occupation number $\langle \alpha|\hat{f}_{k}|\alpha\rangle$ between eigenstates decrease with increasing interaction strength $V$, indicating eigenstate thermalization. Plot shows the variance $ \delta f_{k=0}^{2}$, for states with an effective temperature $T/t=1.2,1.7,\infty$ (from top to bottom), with energy shells of width $\delta E/t=0.25$. Solid lines show ${\rm const}\times\bar{f}_{k=0}(1-\bar{f}_{k=0})\sum_{i}{\rm Var}_{E}(p_{i}^{\alpha})$, with a slightly $T$-dependent constant. Finally, we averaged the results over all total momentum sectors $K$. Inset: As in main figure, but with $\delta f_k^2$ averaged over all $k$. The red dots show $\delta f_k^2$ averaged over all $k$ for an integrable model with nearest-neighbor interactions (at $T=\infty$ and $K/(2\pi/N)=1$).  $K$-averages are only performed to improve statistics. The same results are obtained  for individual $K$-sectors. \label{Fig2}}
\end{figure}

\emph{Fluctuations and IPR}. \textendash{} In the following, we discuss the expectation values 
$f_{k}^{\alpha}$ of the momentum
occupation numbers $\hat{f}_{k}=\hat{c}_{k}^{\dagger}\hat{c}_{k}$ (where 
$\hat{c}_{k}\equiv1/\sqrt{N}\sum_{j=1}^{N}e^{-ikx_{j}}\hat{c}_{j}$).
Being interested in the properties of typical eigenstates, we analyze
the statistics of an ensemble of states $|\alpha\rangle$ with similar
eigenenergies $E_{\alpha}\in I_{E}=[E-\delta E,E+\delta E]$,
which will be called in the following 'eigenstate ensemble' (EE). The width of the energy window $\delta E$ has to be chosen small
enough to avoid artifacts resulting from systematic dependencies
on $E$. Averages with respect to the EE are denoted by $\langle\dots\rangle_{E}$.
For not too large interactions, one can easily show that $\langle f_{k}^{\alpha}\rangle_{E}\approx f_{k,{\rm micro}}(E)$. 
However, the crucial statement of the ETH is that for \emph{each }eigenstate itself $f_{k}^{\alpha}\rightarrow f_{k,{\rm micro}}$
when going to the TDL, i.e., that the fluctuations of $f_{k}^{\alpha}$ from state to state vanish: \begin{eqnarray}
\delta f_{k}^{2}\equiv\left\langle \left\{ f_{k}^{\alpha}-\overline{f_{k}}\right\} ^{2}\right\rangle _{E} & 
\overset{N\rightarrow\infty}{\rightarrow} & 0.\label{eq:f_a_var}\end{eqnarray}
We introduced the EE-variance $\delta f_{k}^{2}$ and $\overline{f_{k}}=\langle f_{k}^{\alpha}\rangle_{E}$. Representing $f_{k}^{\alpha}$ in 
the Fock basis $f_{k}^{\alpha}=\sum_{i=1}^{D_{K}}p_{i}^{\alpha} f_{k}^{i}$
(with $f_{k}^{i}=\langle i|\hat{f}_{k}|i\rangle$) this statement
becomes plausible. For strong interaction, typical
eigenstates are spread out widely in Fock space (Fig.$\,$\ref{Fig1}b), i.e., they are composed of
a large number of Fock states close in energy. Due to the law of large numbers, we thus expect the 
fluctuations to decay as the  mean inverse number of Fock states contributing to $|\alpha\rangle$,
i.e., as the mean IPR \begin{eqnarray}
\chi & = & \langle\sum_{i=1}^{D_{K}}(p_{i}^{\alpha})^{2}\rangle_{E}.\label{eq:}\end{eqnarray}
Before deriving the connection between $\delta f_{k}^{2}$ and
$\chi$ formally, we focus on the numerical results for the present
model. Fig.$\,$\ref{Fig2} shows $\delta f_{k}^{2}$ as a function of $V$
evaluated w.r.t. eigenstates at various energies. The eigenenergies can be re-expressed in terms of effective temperatures $T$, with $E_T\equiv {\rm tr}_K(\hat{H}e^{-\hat{H}/T})/{\rm tr}_K(e^{-\hat{H}/T})$. The
results are compared to the IPR, or more precisely to the sum over the variances ${\rm Var}_E(p_i^{\alpha})=\langle (p_i^{\alpha})^2\rangle_E-\langle p_i^{\alpha}\rangle_E^2$ (see discussion below), clearly demonstrating that indeed $\delta f_{k}^{2}\propto\sum_i {\rm Var}_E(p_i^{\alpha})$
even for small interactions. This is in stark contrast to the case
of integrability conserving nearest-neighbor interaction (inset Fig.$\,$\ref{Fig2}),
where the suppression of $\delta f_{k}^{2}$
with $V$ is much smaller than in the prior case. 

Formally, representing $\delta f_k^2$ in terms of $p_i^{\alpha}$, one finds
\begin{eqnarray}
\delta f_{k}^{2} & \simeq & \overline{f_{k}}(1-\overline{f_{k}})\sum_i {\rm Var}_E(p_i^{\alpha})+\sum_{i\neq j}^{D_{K}}\delta f_{k}^{ij}{\rm Cov}_E(p_i^{\alpha}p_j^{\alpha}),
 \label{eq:1}\end{eqnarray}
with  $\delta f_k^{ij}=(f_{k}^{i}f_{k}^{j}-\overline{f_{k}}^{2})$ and the covariance matrix  ${\rm Cov}_E(p_i^{\alpha} p_j^{\alpha}) \equiv \langle p_i^{\alpha}p_j^{\alpha}\rangle_{E}-\langle p_i^{\alpha}\rangle_E\langle p_j^{\alpha}\rangle_{E}$. The first term in Eq.$\,$(\ref{eq:1}) contains the suppression
of $\delta f_{k}^2$ with increasing number of Fock states contributing
to a typical eigenstate. It is essentially determined by $\chi$ [we note $\chi\approx \sum_i  {\rm Var}_E(p_i^{\alpha})$ below the regime of full chaos (see below)].  We replaced $\sum_i(f_k^i-\overline{f}_k^2) {\rm Var}_E (p_i^{\alpha})$ $\rightarrow  (\overline{f}_k-\overline{f}_k^2)\sum_i{\rm Var}_E(p_i^{\alpha}) $, which is justified as ${\rm Var}_E(p_i^{\alpha})$ is a smooth function of $i$. The prefactor $\overline{f_{k}}(1-\overline{f_{k}})$ is nothing but the variance of the momentum occupation numbers for the non-interacting case.

The off-diagonal contributions in Eq.$\,$(\ref{eq:1}) are sensitive to residual correlations within eigenstates and are expected to become small for strong perturbations. Surprisingly, for strong enough interactions, it approximately reproduces the diagonal part of Eq.$\,$(\ref{eq:1}). Thus, even though $\delta f_k^2$ is still determined by the IPR, one observes a deviation of the prefactor of $\mathcal{O}(1)$. A very similar observation was made in~\cite{1996_FlambaumCorrelations} while investigating finite fermionic systems with random two-body interactions and was traced back to the strong correlations between matrix elements of two-body interaction matrices. 

To sum up, we find that the fluctuations in the expectation value of $\hat{f}_k$ from eigenstate to eigenstate are determined by the IPR $\chi$. Thus, in the following, it will be discussed how $\chi$ decreases with increasing $V$ and system size. Being a measure for the mean
effective number of Fock states forming an eigenstate, $\chi$ indicates
the 'delocalization' crossover in Fock space and serves as an 
indicator for the transition from integrability to quantum chaos. 
  
\begin{figure}
\includegraphics[width=1\columnwidth]{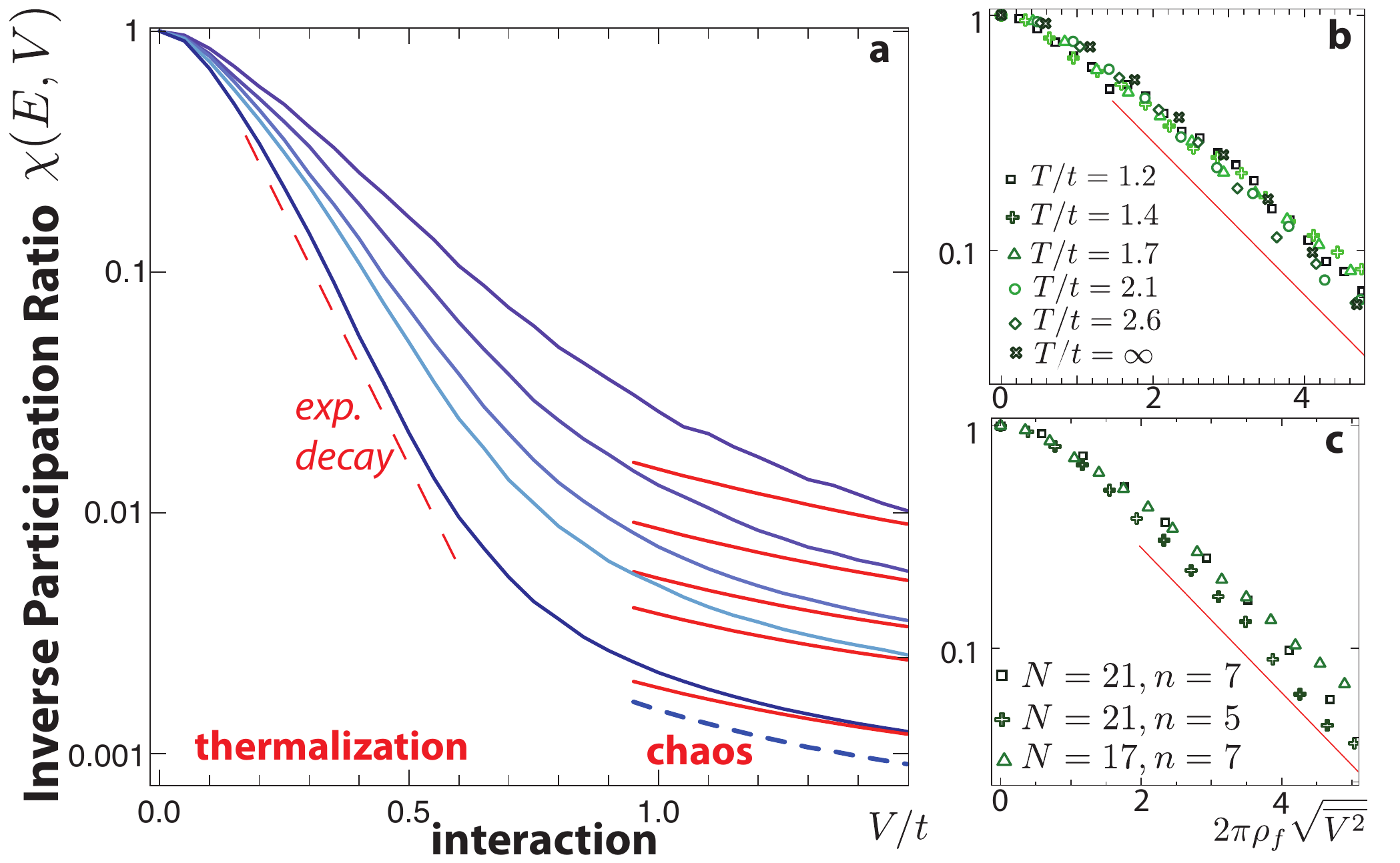}

\caption{a) The inverse participation ratio $\chi$ decreases as many-body eigenstates get more delocalized for increasing interaction strength $V$. From top to bottom: Effective temperatures $T/t=1.2,1.4,1.7,2.6,\infty$ (blue lines; averaged over all momentum sectors). The most important feature is an exponential decay at intermediate interactions (dashed line; see Eq.$\,$(\ref{eq:3})) followed by a power-law tail in the quantum chaotic regime, where it can be well approximated by Eq.$\,$(\ref{eq:6}) (shown only for $T=\infty$ and $K=2\pi/N$: blue, dashed line). In the chaotic regime,  the amplitudes  $c_i^{\alpha}$ are Gaussian distributed, leading to $\chi \rightarrow 3\sum_i\langle p_i^{\alpha}\rangle_E^2$ (red lines).  b) "Scaling plot": As before, but plotted vs. $2\pi\rho_{f}\sqrt{\overline{V^{2}}}$ (and only for a single $K$).  c) Similar plot, but for various system sizes, at $T/t=\infty$.The red line displays Eq.$\,$(\ref{eq:3}), with $\mathcal{C}\approx0.75$, for comparison. }
\label{Fig3}
\end{figure}

 \emph{Definitions} \textendash{} For the following discussion of the IPR, we need to set up a few technical definitions. We introduce the effective density $\rho_{f}^{i}(\omega)$ of Fock states $|j\rangle$ coupling to a state $|i\rangle$
of energy $\xi_{i}\in I_{E}$ (i.e., $\langle i|\hat{V}|j\rangle\neq0$),
where the energy difference between both states is $\xi_{i}-\xi_{j}=\omega$. Averaging over
a couple of states $|i\rangle$ (indicated by $\langle\dots\rangle_{E}^{0}=[\sum_{i,\xi_{i}\in I_{E}}]^{-1}[\sum_{i,\xi_{i}\in I_{E}}\dots]$)
one obtains the \emph{mean effective density of states} $\rho_{f}(\omega,E)=\langle\rho_{f}^{i}(\omega)\rangle_{E}^{0}$.
Furthermore, it will be convenient to introduce the interaction formfactor 
\begin{eqnarray}
F(\omega,E) & = & \pi\langle\int_{\omega-\frac{\delta\omega}{2}}^{\omega+\frac{\delta\omega}{2}}\frac{d\omega'}{\delta\omega}\sum_{j=1;i\neq j}^{D_K}V_{ij}^{2}\delta(\xi_{j}-\xi_{i}-\omega')\rangle_{E}^{0}.\label{eq:2}
\end{eqnarray}
This can be rewritten as $F=\pi \rho_{f}\overline{V^{2}}$,
where $\overline{V^{2}}$ denotes a mean matrix element squared. For $\omega\rightarrow0$
and small $V$, the form factor $F$ reduces to Fermi's golden rule rate for a
Fock state of energy $E$. In the following, only the mean matrix element and the effective density of states 
with respect to states close in energy, i.e., $\overline{V^{2}}(\omega\simeq0,E)$ and $\rho_{f}(\omega\simeq0,E)$ will appear. 
For brevity these will now be denoted by  $\overline{V^{2}}$ and $\rho_{f}=\Delta_f^{-1}$, respectively.

\emph{Localized regime} \textendash{} As long as $\sqrt{\overline{V^{2}}}\ll\rho_{f}^{-1}$,
eigenstates can be obtained within standard perturbation theory (apart from a small set of eigenstates, which can be traced
 back to degenerate Fock states). A given Fock state gets perturbed by the set of directly coupling states
and  eigenstates consist of a small number of sharp peaks (Fig.$\,$\ref{Fig1}a),
i.e., they are localized in Fock space.\\ 
\emph{Delocalization} \textendash{} Increasing the coupling strength $\sqrt{\overline{V^{2}}}\sim\rho_{f}^{-1}$,
one enters the regime of \emph{delocalized} eigenstates~\cite{1997_Georgeot}. Perturbation theory
breaks down and the IPR starts to decrease rapidly (see Fig.$\,$\ref{Fig3}a). In this regime,
the fluctuations $\delta f_{k}^{2}$ become directly determined by
$\chi$. Surprisingly, one observes an exponential decay of $\chi$
and we found good numerical evidence that \begin{eqnarray}\chi & \propto & 
\exp\{ -\mathcal{C}\rho_{f}\sqrt{\overline{V^2}} \}. \label{eq:3}\end{eqnarray}
The numerical constant $\mathcal{C}$ is independent of temperature and system size. In Figs.$\,$\ref{Fig3}b,c, the IPR is shown as a 
function of the scaling variable $2\pi \rho_{f}\sqrt{\overline{V^2}}$ for eigenstates at different energies $E$ and for various $N$ and $n$. 
Indeed, in good accordance to Eq.$\,$(\ref{eq:3}) all curves collapse to the same scaling curve. 
An explanation of this exponential decay of $\chi$ might be found in the two-particle
nature of the interaction, following P.~Silvestrov. In \cite{1998_SilvestrovExp} it was argued (in a random matrix setting) that for moderate interaction strength, typical eigenstates are 
composed of independent pairs of interacting fermions. Thus, eigenstates 
decompose into direct products of pairs of Fock states, resulting in an exponential decay of $\chi$, of the form given by Eq.$\,$($\ref{eq:3}$). While this exponential decay (and additional corrections) have been confirmed numerically in a random quantum dot Hamiltonian~\cite{2000_SilvestrovNum}, here we find it in a translationally invariant many-body system without disorder.

We now discuss the dependence on system size. The effective density of states $\rho_f(\omega)$  scales as $N^3$.  For example, at large $T$, we have $\rho_{f}(\omega)t\simeq N^3 \rho^2(1-\rho)^{2}r(\omega)$, with the density $\rho$. For our particular model, $r(\omega)\propto \ln(t/\omega)$ for $\omega\rightarrow 0$ due to transitions of particle pairs around the inflection point of the $-2t\cos(k)$ dispersion, resulting in $\rho_f^{-1}\propto t/(N^3\ln N)$ (assuming a cutoff scale $\omega/t \sim 1/N$).
Together with the scaling of the matrix elements $\sqrt{\overline{V^{2}}}(\omega,E)=v(\omega,E)V/N$, this would yield $\chi\propto\exp\left\{ -\tilde{\mathcal{C}} N^2 \ln N \rho^2(1-\rho)^{2}V/t\right\} $,
with $\tilde{\mathcal{C}}$ being independent of $N$. Thus, we expect the fluctuations to decrease drastically in the 
thermodynamic limit even in this intermediate regime, where eigenstates are not yet ergodic.  \\ 
\begin{figure}
\includegraphics[width=1\columnwidth]{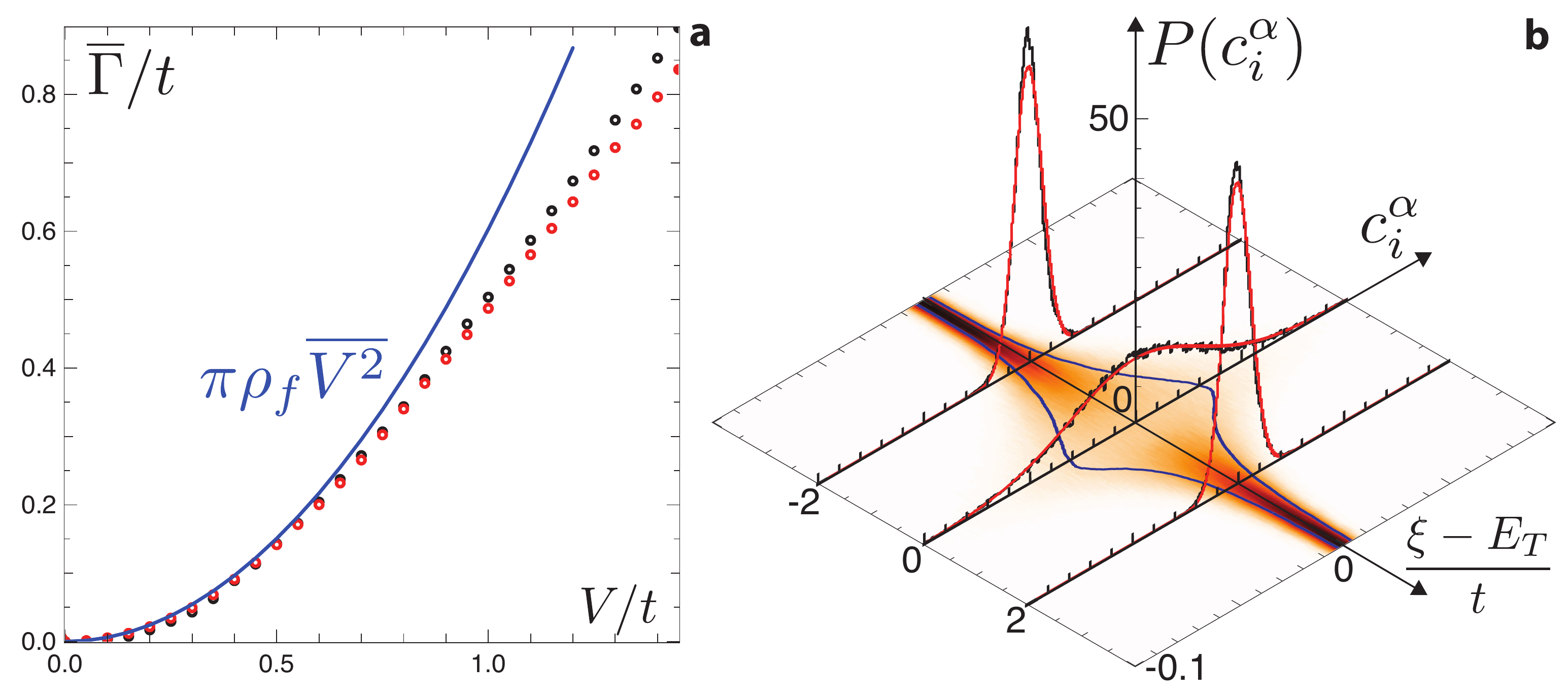}
\caption{a) Fock state decay rate $\overline{\Gamma}$ vs. $V$ (at $T=\infty$), extracted from the imaginary part of the self-energy (red dots). For small $V$, Fermi's golden rule $\overline{\Gamma}=\pi\rho_{f}\overline{V^{2}}$ holds, while $\overline{\Gamma}$ increases linearly in $V$ for large $V$. For this plot, ${\rm Im}\Sigma_{i}(\omega)$ was averaged in both $\omega$ and energy $\xi_{i}$ over the energy interval of width $\delta E=0.25t$ centered around $E_T$. Black dots show the results of a direct fit of $\langle p_i^{\alpha}\rangle_E$. Here $K=2\pi/N$. b) Amplitude distribution $P(c_i^{\alpha})$  for eigenstates at $T/t=\infty$  with $\delta E/t=0.25$ and $V/t=1$ demonstrating that for very large $V$ one enters the chaotic regime. In this regime, the amplitudes $c_i^{\alpha}$ are gaussian distributed as originally conjectured in ~\cite{1991_Deutsch}. In plane: energy dependent standard deviation $\pm[\langle (c_i^{\alpha})^2\rangle_E]^{1/2}$ of $P$ (blue lines). Out of plane: Cuts of $P$ (black lines), which can be described by gaussians of variance $\langle (c_i^{\alpha})^2\rangle_E$ (red lines). }
\label{Fig4}
\end{figure}
\emph{Chaos} \textendash{} Only by increasing the interaction even further, one enters the regime of ergodic eigenstates. 
By 'ergodic eigenstates', we understand states which in principle are composed of all 
Fock states close in energy (cf. Fig.$\,$\ref{Fig1}b). No Fock states are excluded a priori, e.g., due to 
the two-body nature of $\hat{V}$ or further symmetries from contributing to an ergodic 
eigenstate. The amplitudes $c_i^{\alpha}$ become Gaussian distributed random 
variables~\cite{1991_Deutsch,1994_Srednicki} as it is shown in Fig.$\,$\ref{Fig4}b with a Lorentzian variance~\cite{1991_Deutsch, 1994_Lauritzen, 2000_Flambaum}
\begin{eqnarray} \langle p_i^{\alpha}\rangle_E & \simeq  & \frac{1}{\pi \rho_K(E)}\frac{\overline{\Gamma}(E-\xi_i,E)}{(\xi_i-E-\overline{\delta}(E,\xi_i))^2
+\overline{\Gamma}^2} ,\label{eq:4}\end{eqnarray} 
where $\rho_K$ denotes the full many-body density of states for total momentum $K$, scaling as $(N-1)!/(N-n)!n! $. 
This indicates the crossover to full quantum chaos. We checked that in this regime the nearest 
neighbor level spacing statistics agrees with the  GOE-Wigner surmise, characteristic for GOE random matrix ensembles. 
Due to the Gaussian distribution for
$c_i^{\alpha}$, one finds $\chi= 3\sum_{\alpha}\langle p_i^{\alpha}\rangle_E^2$ resulting in
\begin{eqnarray}  \chi & \simeq & \frac{3}{2\pi}\frac{1}{\overline{\Gamma}(0,E)\rho_K(E)}
 \label{eq:6},\end{eqnarray} 
 which is in fairly good agreement with the numerical results in Fig.$\,$(\ref{Fig3}a), demonstrating the suppression of $\delta f_k^2$ by the inverse many-body density of states as it was conjectured in~\cite{1991_Deutsch,1994_Srednicki}. The mean spreading width $\overline{\Gamma}$ (Fig.$\,$\ref{Fig4}a) can be extracted from the Fock state self-energy $\Sigma$ by averaging $ -\rm{Im}\Sigma(\xi_i,\omega)$ over $\xi_i, \omega \in I_E$. $\Sigma$ is obtained from $G(\xi_i,\omega)\equiv \langle i|[\omega +i0^+-\hat{H}]^{-1} |i\rangle$ via $G\equiv[\omega+i0^+-\xi_i-\Sigma]^{-1}$.  Fig.$\,$\ref{Fig1}c shows a comparison of  $\langle p_i^{\alpha}\rangle_E$ and a Lorentzian of width $\overline{\Gamma}$ extracted directly from $-\rm{Im}\Sigma$.\\
The important question remains, how the second crossover scale (governing the crossover from delocalized to ergodic eigenstates) depends on system size. We found some indication that it might depend on the intensive 'energy range'  $W$ of the coupling matrix $\hat{V}$. Consider the dependence of $\overline{\Gamma}$ on $V$ in Fig.$\,$\ref{Fig4}a. For small $V$, Fermi's golden rule applies and one finds $\overline{\Gamma}(0,E)\simeq  F(0,E)\propto V^2/t$. For large $V$, one observes a crossover $\overline{\Gamma}\propto 
 V^2/t\rightarrow \overline{\Gamma} \propto V$  indicating the entrance into the strong coupling regime, where 
 $\overline{\Gamma}$ and the finite width (in $\omega$) of the formfactor $F$ become comparable~\cite{1994_Lauritzen}. 
 Comparing Figs.$\,$\ref{Fig3}a and \ref{Fig4}a, there might exist a close relation between this crossover and the onset of ergodicity of eigenstates. 
 This would imply that the interaction energy 
 scale $\rho_f^{-1}$ for the onset of thermalization is parametrically smaller than the scale for the transition to chaos, determined by $W$.
 
 \emph{Conclusions}. \textendash{} By means of exact diagonalization we investigated the interaction induced onset of eigenstate thermalization in  a system of 1D fermions. We found that the fluctuations of the expectation value of the  momentum occupation number from  state to state are proportional to the inverse participation ratio of eigenstates. For small interactions the latter decays exponentially before one enters the chaotic regime. The interaction scale for the onset of this decay is essentially set by the effective mean level spacing between interacting Fock states, and this vanishes in the TDL. Thus,  we corroborate the physical expectation that in the TDL at arbitrarily small interactions, eigenstate thermalization sets in.  
 
\emph{Acknowledgements}. \textendash{} We thank P.G. Silvestrov, Frank G\"ohmann and A. Polkovnikov for fruitful discussions related to this work.
Financial support by DIP, NIM, the Emmy-Noether program and the SFB/TR
12 is gratefully acknowledged.
 
\bibliographystyle{apsrev4-1}
\bibliography{/Users/Neuenhahn/Documents/Promotion/BIB/bibPromotion}

\begin{thebibliography}{10}%
\makeatletter
\providecommand \@ifxundefined [1]{%
 \ifx #1\undefined \expandafter \@firstoftwo
 \else \expandafter \@secondoftwo
\fi
}%
\providecommand \@ifnum [1]{%
 \ifnum #1\expandafter \@firstoftwo
 \else \expandafter \@secondoftwo
\fi
}%
\providecommand \enquote [1]{``#1''}%
\providecommand \bibnamefont  [1]{#1}%
\providecommand \bibfnamefont [1]{#1}%
\providecommand \citenamefont [1]{#1}%
\providecommand\href[0]{\@sanitize\@href}%
\providecommand\@href[1]{\endgroup\@@startlink{#1}\endgroup\@@href}%
\providecommand\@@href[1]{#1\@@endlink}%
\providecommand \@sanitize [0]{\begingroup\catcode`\&12\catcode`\#12\relax}%
\@ifxundefined \pdfoutput {\@firstoftwo}{%
 \@ifnum{\z@=\pdfoutput}{\@firstoftwo}{\@secondoftwo}%
}{%
 \providecommand\@@startlink[1]{\leavevmode\special{html:<a href="#1">}}%
 \providecommand\@@endlink[0]{\special{html:</a>}}%
}{%
 \providecommand\@@startlink[1]{%
  \leavevmode
  \pdfstartlink
   attr{/Border[0 0 1 ]/H/I/C[0 1 1]}%
   user{/Subtype/Link/A<</Type/Action/S/URI/URI(#1)>>}%
  \relax
 }%
 \providecommand\@@endlink[0]{\pdfendlink}%
}%
\providecommand \url  [0]{\begingroup\@sanitize \@url }%
\providecommand \@url [1]{\endgroup\@href {#1}{\urlprefix}}%
\providecommand \urlprefix [0]{URL }%
\providecommand \Eprint[0]{\href }%
\@ifxundefined \urlstyle {%
  \providecommand \doi [1]{doi:\discretionary{}{}{}#1}%
}{%
  \providecommand \doi [0]{doi:\discretionary{}{}{}\begingroup
  \urlstyle{rm}\Url }%
}%
\providecommand \doibase [0]{http://dx.doi.org/}%
\providecommand \Doi[1]{\href{\doibase#1}}%
\providecommand \bibAnnote [3]{%
  \BibitemShut{#1}%
  \begin{quotation}\noindent
    \textsc{Key:}\ #2\\\textsc{Annotation:}\ #3%
  \end{quotation}%
}%
\providecommand \bibAnnoteFile [2]{%
  \IfFileExists{#2}{\bibAnnote {#1} {#2} {\input{#2}}}{}%
}%
\providecommand \typeout [0]{\immediate \write \m@ne }%
\providecommand \selectlanguage [0]{\@gobble}%
\providecommand \bibinfo [0]{\@secondoftwo}%
\providecommand \bibfield [0]{\@secondoftwo}%
\providecommand \translation [1]{[#1]}%
\providecommand \BibitemOpen[0]{}%
\providecommand \bibitemStop [0]{}%
\providecommand \bibitemNoStop [0]{.\EOS\space}%
\providecommand \EOS [0]{\spacefactor3000\relax}%
\providecommand \BibitemShut [1]{\csname bibitem#1\endcsname}%
\bibitem{1955_Fermi}%
  \BibitemOpen
  \bibfield{author}{%
  \bibinfo {author} {\bibfnamefont{P.~J.}\ \bibnamefont{Fermi~E.}}\ and\
  \bibinfo {author} {\bibfnamefont{U.}~\bibnamefont{S.}},\ }%
  \bibfield{journal}{%
  \bibinfo {journal} {Los Alamos report LA-1940}}%
   (\bibinfo {year} {1955})%
  \bibAnnoteFile{NoStop}{1955_Fermi}%
\bibitem{1992_Ford_Fermi_Pasta_Ulam}%
  \BibitemOpen
  \bibfield{author}{%
  \bibinfo {author} {\bibfnamefont{J.}~\bibnamefont{Ford}},\ }%
  \bibfield{journal}{%
  \Doi{DOI: 10.1016/0370-1573(92)90116-H}{\bibinfo {journal} {Physics
  Reports}}\ }%
  \textbf{\bibinfo {volume} {213}},\ \bibinfo {pages} {271 } (\bibinfo {year}
  {1992}),\ ISSN \bibinfo {issn} {0370-1573}%
  \bibAnnoteFile{NoStop}{1992_Ford_Fermi_Pasta_Ulam}%
\bibitem{2002_GReinerCollapseRevival}%
  \BibitemOpen
  \bibfield{author}{%
  \bibinfo {author} {\bibfnamefont{M.}~\bibnamefont{Greiner}}, \bibinfo
  {author} {\bibfnamefont{O.}~\bibnamefont{Mandel}}, \bibinfo {author}
  {\bibfnamefont{T.~W.}\ \bibnamefont{Hansch}},\ and\ \bibinfo {author}
  {\bibfnamefont{I.}~\bibnamefont{Bloch}},\ }%
  \bibfield{journal}{%
  \bibinfo {journal} {Nature}\ }%
  \textbf{\bibinfo {volume} {419}},\ \bibinfo {pages} {51} (\bibinfo {month}
  {09}\ \bibinfo {year} {2002})%
  \bibAnnoteFile{NoStop}{2002_GReinerCollapseRevival}%
\bibitem{2006_Kinoshita}%
  \BibitemOpen
  \bibfield{author}{%
  \bibinfo {author} {\bibfnamefont{T.}~\bibnamefont{Kinoshita}}, \bibinfo
  {author} {\bibfnamefont{T.}~\bibnamefont{Wenger}},\ and\ \bibinfo {author}
  {\bibfnamefont{D.~S.}\ \bibnamefont{Weiss}},\ }%
  \bibfield{journal}{%
  \bibinfo {journal} {Nature}\ }%
  \textbf{\bibinfo {volume} {440}},\ \bibinfo {pages} {900} (\bibinfo {month}
  {04}\ \bibinfo {year} {2006})%
  \bibAnnoteFile{NoStop}{2006_Kinoshita}%
\bibitem{2007_Hofferbeth}%
  \BibitemOpen
  \bibfield{author}{%
  \bibinfo {author} {\bibfnamefont{S.}~\bibnamefont{Hofferberth}}, \bibinfo
  {author} {\bibfnamefont{I.}~\bibnamefont{Lesanovsky}}, \bibinfo {author}
  {\bibfnamefont{B.}~\bibnamefont{Fischer}}, \bibinfo {author}
  {\bibfnamefont{T.}~\bibnamefont{Schumm}},\ and\ \bibinfo {author}
  {\bibfnamefont{J.}~\bibnamefont{Schmiedmayer}},\ }%
  \bibfield{journal}{%
  \bibinfo {journal} {Nature}\ }%
  \textbf{\bibinfo {volume} {449}},\ \bibinfo {pages} {324} (\bibinfo {month}
  {09}\ \bibinfo {year} {2007})%
  \bibAnnoteFile{NoStop}{2007_Hofferbeth}%
\bibitem{1991_Deutsch}%
  \BibitemOpen
  \bibfield{author}{%
  \bibinfo {author} {\bibfnamefont{J.~M.}\ \bibnamefont{Deutsch}},\ }%
  \bibfield{journal}{%
  \Doi{10.1103/PhysRevA.43.2046}{\bibinfo {journal} {Phys. Rev. A}}\ }%
  \textbf{\bibinfo {volume} {43}},\ \bibinfo {pages} {2046} (\bibinfo {month}
  {Feb}\ \bibinfo {year} {1991})%
  \bibAnnoteFile{NoStop}{1991_Deutsch}%
\bibitem{1994_Srednicki}%
  \BibitemOpen
  \bibfield{author}{%
  \bibinfo {author} {\bibfnamefont{M.}~\bibnamefont{Srednicki}},\ }%
  \bibfield{journal}{%
  \Doi{10.1103/PhysRevE.50.888}{\bibinfo {journal} {Phys. Rev. E}}\ }%
  \textbf{\bibinfo {volume} {50}},\ \bibinfo {pages} {888} (\bibinfo {month}
  {Aug}\ \bibinfo {year} {1994})%
  \bibAnnoteFile{NoStop}{1994_Srednicki}%
\bibitem{2008_RigolNature}%
  \BibitemOpen
  \bibfield{author}{%
  \bibinfo {author} {\bibfnamefont{M.}~\bibnamefont{Rigol}}, \bibinfo {author}
  {\bibfnamefont{V.}~\bibnamefont{Dunjko}},\ and\ \bibinfo {author}
  {\bibfnamefont{M.}~\bibnamefont{Olshanii}},\ }%
  \bibfield{journal}{%
  \bibinfo {journal} {Nature}\ }%
  \textbf{\bibinfo {volume} {452}},\ \bibinfo {pages} {854} (\bibinfo {year}
  {2008})%
  \bibAnnoteFile{NoStop}{2008_RigolNature}%
\bibitem{2009_RigolBreakdown}%
  \BibitemOpen
  \bibfield{author}{%
  \bibinfo {author} {\bibfnamefont{M.}~\bibnamefont{Rigol}},\ }%
  \bibfield{journal}{%
  \Doi{10.1103/PhysRevLett.103.100403}{\bibinfo {journal} {Phys. Rev. Lett.}}\
  }%
  \textbf{\bibinfo {volume} {103}},\ \bibinfo {pages} {100403} (\bibinfo
  {month} {Sep}\ \bibinfo {year} {2009})%
  \bibAnnoteFile{NoStop}{2009_RigolBreakdown}%
\bibitem{2009_RigolFermions}%
  \BibitemOpen
  \bibfield{author}{%
  \bibinfo {author} {\bibfnamefont{M.}~\bibnamefont{Rigol}},\ }%
  \bibfield{journal}{%
  \Doi{10.1103/PhysRevA.80.053607}{\bibinfo {journal} {Phys. Rev. A}}\ }%
  \textbf{\bibinfo {volume} {80}},\ \bibinfo {pages} {053607} (\bibinfo {month}
  {Nov}\ \bibinfo {year} {2009})%
  \bibAnnoteFile{NoStop}{2009_RigolFermions}%
\bibitem{2010_Santos}%
  \BibitemOpen
  \bibfield{author}{%
  \bibinfo {author} {\bibfnamefont{L.~F.}\ \bibnamefont{Santos}}\ and\ \bibinfo
  {author} {\bibfnamefont{M.}~\bibnamefont{Rigol}},\ }%
  \bibfield{journal}{%
  \Doi{10.1103/PhysRevE.81.036206}{\bibinfo {journal} {Phys. Rev. E}}\ }%
  \textbf{\bibinfo {volume} {81}},\ \bibinfo {pages} {036206} (\bibinfo {month}
  {Mar}\ \bibinfo {year} {2010})%
  \bibAnnoteFile{NoStop}{2010_Santos}%
\bibitem{2010_Canovi}%
  \BibitemOpen
  \bibfield{author}{%
  \bibinfo {author} {\bibfnamefont{E.}~\bibnamefont{Canovi}}, \bibinfo {author}
  {\bibfnamefont{D.}~\bibnamefont{Rossini}}, \bibinfo {author}
  {\bibfnamefont{R.}~\bibnamefont{Fazio}}, \bibinfo {author}
  {\bibfnamefont{G.~E.}\ \bibnamefont{Santoro}},\ and\ \bibinfo {author}
  {\bibfnamefont{A.}~\bibnamefont{Silva}},\ }%
  \bibfield{journal}{%
  \bibinfo {journal} {arxiv:1006.1634v1}}%
   (\bibinfo {year} {2010})%
  \bibAnnoteFile{NoStop}{2010_Canovi}%
\bibitem{2009_Cassidy}%
  \BibitemOpen
  \bibfield{author}{%
  \bibinfo {author} {\bibfnamefont{A.~C.}\ \bibnamefont{Cassidy}}, \bibinfo
  {author} {\bibfnamefont{D.}~\bibnamefont{Mason}}, \bibinfo {author}
  {\bibfnamefont{V.}~\bibnamefont{Dunjko}},\ and\ \bibinfo {author}
  {\bibfnamefont{M.}~\bibnamefont{Olshanii}},\ }%
  \bibfield{journal}{%
  \Doi{10.1103/PhysRevLett.102.025302}{\bibinfo {journal} {Physical Review
  Letters}}\ }%
  \textbf{\bibinfo {volume} {102}},\ \bibinfo {eid} {025302} (\bibinfo {year}
  {2009})%
  \bibAnnoteFile{NoStop}{2009_Cassidy}%
\bibitem{2000_SilvestrovNum}%
  \BibitemOpen
  \bibfield{author}{%
  \bibinfo {author} {\bibfnamefont{X.}~\bibnamefont{Leyronas}}, \bibinfo
  {author} {\bibfnamefont{P.~G.}\ \bibnamefont{Silvestrov}},\ and\ \bibinfo
  {author} {\bibfnamefont{C.~W.~J.}\ \bibnamefont{Beenakker}},\ }%
  \bibfield{journal}{%
  \Doi{10.1103/PhysRevLett.84.3414}{\bibinfo {journal} {Phys. Rev. Lett.}}\ }%
  \textbf{\bibinfo {volume} {84}},\ \bibinfo {pages} {3414} (\bibinfo {month}
  {Apr}\ \bibinfo {year} {2000})%
  \bibAnnoteFile{NoStop}{2000_SilvestrovNum}%
\bibitem{1997_Silvestrov}%
  \BibitemOpen
  \bibfield{author}{%
  \bibinfo {author} {\bibnamefont{P.G.Silvestrov}},\ }%
  \bibfield{journal}{%
  \bibinfo {journal} {Phys Rev Lett}\ }%
  \textbf{\bibinfo {volume} {79}} (\bibinfo {year} {1997})%
  \bibAnnoteFile{NoStop}{1997_Silvestrov}%
\bibitem{2009_Olshanii}%
  \BibitemOpen
  \bibfield{author}{%
  \bibinfo {author} {\bibfnamefont{M.}~\bibnamefont{Olshanii}}\ and\ \bibinfo
  {author} {\bibfnamefont{V.}~\bibnamefont{Yurovsky}},\ }%
  \bibfield{journal}{%
  \bibinfo {journal} {arxiv: 0911.5587v1}}%
   (\bibinfo {year} {2009})%
  \bibAnnoteFile{NoStop}{2009_Olshanii}%
\bibitem{1997_AlthsulerLoc}%
  \BibitemOpen
  \bibfield{author}{%
  \bibinfo {author} {\bibfnamefont{B.~L.}\ \bibnamefont{Altshuler}}, \bibinfo
  {author} {\bibfnamefont{Y.}~\bibnamefont{Gefen}}, \bibinfo {author}
  {\bibfnamefont{A.}~\bibnamefont{Kamenev}},\ and\ \bibinfo {author}
  {\bibfnamefont{L.~S.}\ \bibnamefont{Levitov}},\ }%
  \bibfield{journal}{%
  \Doi{10.1103/PhysRevLett.78.2803}{\bibinfo {journal} {Phys. Rev. Lett.}}\ }%
  \textbf{\bibinfo {volume} {78}},\ \bibinfo {pages} {2803} (\bibinfo {month}
  {Apr}\ \bibinfo {year} {1997})%
  \bibAnnoteFile{NoStop}{1997_AlthsulerLoc}%
\bibitem{1997_Georgeot}%
  \BibitemOpen
  \bibfield{author}{%
  \bibinfo {author} {\bibfnamefont{B.}~\bibnamefont{Georgeot}}\ and\ \bibinfo
  {author} {\bibfnamefont{D.~L.}\ \bibnamefont{Shepelyansky}},\ }%
  \bibfield{journal}{%
  \Doi{10.1103/PhysRevLett.79.4365}{\bibinfo {journal} {Phys. Rev. Lett.}}\ }%
  \textbf{\bibinfo {volume} {79}},\ \bibinfo {pages} {4365} (\bibinfo {month}
  {Dec}\ \bibinfo {year} {1997})%
  \bibAnnoteFile{NoStop}{1997_Georgeot}%
\bibitem{1996_FlambaumCorrelations}%
  \BibitemOpen
  \bibfield{author}{%
  \bibinfo {author} {\bibfnamefont{V.~V.}\ \bibnamefont{Flambaum}}, \bibinfo
  {author} {\bibfnamefont{G.~F.}\ \bibnamefont{Gribakin}},\ and\ \bibinfo
  {author} {\bibfnamefont{F.~M.}\ \bibnamefont{Izrailev}},\ }%
  \bibfield{journal}{%
  \Doi{10.1103/PhysRevE.53.5729}{\bibinfo {journal} {Phys. Rev. E}}\ }%
  \textbf{\bibinfo {volume} {53}},\ \bibinfo {pages} {5729} (\bibinfo {month}
  {Jun}\ \bibinfo {year} {1996})%
  \bibAnnoteFile{NoStop}{1996_FlambaumCorrelations}%
\bibitem{1998_SilvestrovExp}%
  \BibitemOpen
  \bibfield{author}{%
  \bibinfo {author} {\bibfnamefont{P.~G.}\ \bibnamefont{Silvestrov}},\ }%
  \bibfield{journal}{%
  \Doi{10.1103/PhysRevE.58.5629}{\bibinfo {journal} {Phys. Rev. E}}\ }%
  \textbf{\bibinfo {volume} {58}},\ \bibinfo {pages} {5629} (\bibinfo {month}
  {Nov}\ \bibinfo {year} {1998})%
  \bibAnnoteFile{NoStop}{1998_SilvestrovExp}%
\bibitem{1994_Lauritzen}%
  \BibitemOpen
  \bibfield{author}{%
  \bibinfo {author} {\bibfnamefont{B.}~\bibnamefont{Lauritzen}}\ and\ \bibinfo
  {author} {\bibfnamefont{V.}~\bibnamefont{Zelevinsky}},\ }%
  \bibfield{journal}{%
  \bibinfo {journal} {Phys. Rev. Let.}\ }%
  \textbf{\bibinfo {volume} {74}},\ \bibinfo {pages} {5190} (\bibinfo {year}
  {1995})%
  \bibAnnoteFile{NoStop}{1994_Lauritzen}%
\bibitem{2000_Flambaum}%
  \BibitemOpen
  \bibfield{author}{%
  \bibinfo {author} {\bibfnamefont{V.~V.}\ \bibnamefont{Flambaum}}\ and\
  \bibinfo {author} {\bibfnamefont{F.~M.}\ \bibnamefont{Izrailev}},\ }%
  \bibfield{journal}{%
  \bibinfo {journal} {Phys. Rev. E}\ }%
  \textbf{\bibinfo {volume} {61}} (\bibinfo {year} {2000})%
  \bibAnnoteFile{NoStop}{2000_Flambaum}%
\end{thebibliography}%

\end{document}